\documentstyle[eqsecnum,aps,epsf]{revtex}

\begin{document}
\preprint{HEP/123-qed}
\title{Anomalous jumping
in a double-well potential
}
\author{P. D. Ditlevsen\\
The Niels Bohr Institute, Department for Geophysics,\\
University of Copenhagen, Juliane Maries Vej 30,\\
 DK-2100 Copenhagen O, Denmark.}
\date{\today}
\maketitle
\begin{abstract}
Noise induced jumping between meta-stable states in a potential depends
on the structure of the noise. For an $\alpha$-stable noise, jumping
triggered by single extreme events contributes to the transition
probability. This is also called Levy flights and
might be of importance in triggering sudden changes in geophysical
flow and perhaps even climatic changes. The steady state statistics
is also influenced by the noise structure leading to a non-Gibbs
distribution for an $\alpha$-stable noise.
\end{abstract}
\pacs{PACS appear here.}

\section{Introduction}
Noise induced jumping between meta-stable states separated by potential
barriers is common in
physical systems.
The time scale for
the barrier penetration depends on the 
structure of the noise.
Most often the noise is Gaussian. However, non-Gaussian 
noises distributed with power-function tails, Levy flights, are observed in
many different physical systems \cite{Frisch} such as
turbulent diffusion \cite{Shlesinger,Zimbardo}  and vortex 
dynamics \cite{Viecelli}. 
Levy flights also seems to be a common feature in
dynamical models \cite{Klafter} and critical phenomena \cite{Bak}.

The Levy flights can result from a Langevin equation driven by
$\alpha$-stable noise  and  give rise 
to anomalous diffusion of a 
random walker with position $r(t)$ such that 
$\langle |r(t)-r(0)|^2 \rangle \propto D t^{2/\alpha}$
where $D$ is a constant and
$0 < \alpha < 2$ \cite{bou}. The case $\alpha =2$ corresponds to normal diffusion
where $D$ is the diffusion constant. 
The exponent
$\alpha$ is related to the scaling of the tail of the probability
distribution for the increments of the random walker, 
$P(X>r)\propto r^{-\alpha}$. For
$\alpha \ge 2$ the second moment exists and by the central
limit theorem the random walker reduces in the continuum 
limit to a Gaussian
random walker unless the diffusion takes place on a fractal
set like in a quenched random medium \cite{Fogedby}. In this
case the random walk can be sub-diffusive. Another example
of a process which can be sub-diffusive is the Levy walk where a
random walker has a constant speed in between discrete 
stochastic time points (a renewal process) with a power-function tail distribution.
Note that since the time process is discrete for a Levy walk it 
cannot result from a Langevin
equation.

Anomalous diffusion was first observed in hydrological
time-series \cite{Hurst}.
Recently evidence for $\alpha$-stable statistics in atmospheric 
circulation data has been reported \cite{Viecelli1}. In a long
paleoclimatic time-series
an $\alpha$-stable noise induced jumping in a double-well potential was
found \cite{Ditlevsen}. In both cases $\alpha$ was found to be around 1.7.
The latter describes a jumping, in glacial times, between two
climatic states governed by the oceanic flow forced by random
fluctuations from the atmosphere.
Understanding the role of extreme events and the time-scales 
for these climatic shifts
is the main motivation for this study.

In this paper we will interchangeably use the physics
jargon, $\langle x\rangle$, and the mathematics jargon,
$E[x]$, for the expectation value for $x$. The latter will
be used in the case of conditional expectations.
We use the usual convention that probability
distribution functions, $P$, are capitalized and probability 
density functions, $p=dP/dx$, are in small letters.

\section{The $\alpha$-stable distributions}
For distributions with power-function tails, $P(X>x)\propto x^{-\gamma}$,
only moments of order less than $\gamma$
exists $(\langle |x|^{\beta}\rangle = \infty \, \mbox{for} \, \beta
\geq \gamma)$. For $0<\gamma<2$ a generalized version of the
central limit theorem applies, namely that the average of $n$ independent stochastic variables 
from the distribution $P$ asymptotically will have an $\alpha$-stable distribution
as $n\rightarrow\infty$ with $\alpha=\gamma$. 
The $\alpha$-stable distributions are 
defined by their characteristic functions, 
$\langle \exp(ikX)\rangle =\exp(-\sigma^\alpha 
|k|^\alpha/\alpha)$.
The
$\alpha$-stable distributions are stable with respect to averaging,
$Y_n =n^{-1/\alpha}\sum_{i=1}^n X_i$, meaning that $Y_n$ has the
same distribution as $X_i$ where the $X_i$'s are i.i.d. (independent
identically distributed) $\alpha$-stable,
thus the phrase '$\alpha$-stable'. 
As for the case of Gaussian noise, the dynamics of a noise driven 
system
with power-function tail distributions for the noise increments,
$P(X>x)\propto x^{-\alpha}, 0<\alpha<2$, 
will reduce to a system with an $\alpha$-stable noise
in the continuum limit, described by a Langevin equation \cite{Protter},

\begin{equation}
dX=f(X)dt + \sigma(X) dL_{\alpha}.
\label{LE}
\end{equation}

A random walker with $\alpha$-stable
noise increments will be super-diffusive due to the large jumps from
the tails of the distribution surviving the averaging in the
continuum limit. See Appendix A for a further short description.

\section{The Fokker-Planck equation}
The probability density for $X$ in (\ref{LE}) is determined from
the Fokker-Planck equation (FPE), see Appendix B for a derivation,

\begin{eqnarray}
\partial_t p(x) = -\partial_x
[f(x)p(x)]
\nonumber \\
-\frac{1}{\alpha}\int\int e^{-ikx} \widehat{\sigma^{\alpha}}(k-k_1) 
|k|^{\alpha} \widehat{ p}(k_1) dk dk_1 .
\label{FP}
\end{eqnarray}
The second term on the right hand side is expressed 
in terms of the
Fourier transformed probability density, $\widehat{ p}(k)$. This term reduces to 
the ordinary diffusion term, $\partial_{x^2} [\sigma^2(x)
p(x)]/2$.
when $\alpha=2$. 
In this case the solution for the stationary probability
density function can be expressed explicitly in 
the well-known form,

\begin{equation}
p(x)\propto \frac{1}{\sigma^2(x)}\exp\{2\int_0^x 
\frac{f(y)}{\sigma^2(y)}dy\},
\label{pdf}
\end{equation}
For $\alpha < 2$ the FPE (\ref{FP}) is non-local in spectral
space. This is a reflection of the super-diffusivity of the process (\ref{LE}).
Besides the Gaussian case we can only solve the FPE
explicitly for $\alpha=1$. This is the case of a system
driven by Cauchy distributed noise having the
probability density, 
$q(x)=1/[\pi(1+x^2)]$, see Appendix C for further details on the
Cauchy distribution. We are using $p(x)$ for the probability density
for $X$ in (\ref{LE}) and $q(x)$ for the probability density
of the noise.
Then the stationary FPE becomes, 
\begin{equation}
i\int \widehat{f}(k_1-k)\widehat{p}(k_1)dk_1=\mbox{sgn}(k)\sigma
\widehat{p}(k)
\label{Cauchy1}
\end{equation}
where the noise intensity $\sigma$ is taken to be constant.
From taking the derivative with respect to $k$ on both sides of
(\ref{Cauchy1}) and performing
a partial integration on the left-hand side it follows that
 
\begin{equation}
i\int \widehat{f}(k_1-k)\widehat{p}^{(m)}(k_1)dk_1=\mbox{sgn}(k)\sigma
\widehat{p}^{(m)}(k)
\label{Cauchy2}
\end{equation}
for any $m$. The solution is $\widehat{p}(k)=e^{-\lambda |k|}$,
where $\lambda $ is determined by
 
\begin{equation}
i\int \widehat{f}(k_1-k)e^{-\lambda (k_1-k)}dk_1 =
if(i\lambda ) =\mbox{sgn}(k)\sigma. \label{Feq}
\label{Cauchy3}
\end{equation}
Thus the solution is determined by the analytic continuation of $f(x)$
into the
complex plane, provided it exists.
Note that the solution also apply for $k=0$ where the r.h.s of (\ref{Cauchy2})
jumps, since from the definition of the Fourier transform of the probability density we have $\hat{p}(0)=E[1]=1$.   
By complex conjugation of (\ref{Cauchy3})
we get
$if(-i\lambda ^{*})=\mbox{sgn}(-k)\sigma$,
so for $k<0$ the solution is given as $-\lambda ^{*}$, where $\lambda $
solves
(\ref{Cauchy3}) for $k>0$.  With $\lambda =\beta +i\delta $ the
characteristic
function is given as $\widehat{p}(k)=e^{-\beta |k|}e^{-i\delta k}$. For
$%
\widehat{p}(k)$ to be a characteristic function we must have
$\beta >0,
$ and the stationary distribution is 
\begin{equation}
p(x)=\sum_{i=1}^N p_i\frac 1\pi \frac{\beta _i}{\beta _i^2+(x+\delta _i)^2}
\label{8}
\end{equation}
where the
sum is
over the $N$ zero points of the complex function $if(i\lambda
)-\mbox{sgn}(k)\sigma $
in the upper half-plane  $(\beta >0)$. In this solution of the 
stationary FPE 
there is
an indeterminacy since any $p(x)$ with $\sum_ip_i =1$ is a probability
density that satisfies (\ref{FP}).

The indeterminacy might be related to the problem of conservation 
of probability. If there is a finite probability for the random
walker to escape to infinity, it must be reinserted into the 
system for a stationary probability density to be conserved. Then the indeterminacy in
the reinsertion could result in the indeterminacy in the coefficients
$p_i$ in (\ref{8}). However, the indeterminacy can be lifted in
the limit $\sigma \rightarrow 0$. When the intensity of the noise
becomes small the Cauchy distribution approaches a $\delta$-distribution (when
acting on functions that are bounded by $|x|^\beta$ for some $\beta<\alpha$ as
$x\rightarrow \infty$).
Then we can approximate the system by a system with discrete states and the
stationary Fokker-Planck equation (\ref{Cauchy1}) is approximated by $N$ transition (Master) equations for the weights $p_i, i=1,...N$,

\begin{equation}
p_i = \Sigma_j p_j p(j \rightarrow i),
\label{Master}
\end{equation}
where $i, j$ represents the $N$ minima defined in (\ref{8}). The transition probabilities $ p(i \rightarrow j)$ are related to the 
transition waiting times which will be defined in the following.
 
\section{The potential}
Before proceeding we will define the drift term as
resulting from a potential.
The governing equation then describes a massless, viscous particle 
moving in a potential, $f(x)=-dU/dx$. As an example for study 
we define the potential as

\begin{equation}
U(x)=4(x/\Delta)^4+h(x/\Delta)^3-8(x/\Delta)^2-3h(x/\Delta).
\label{potential}
\end{equation}
$U(x)$
is a double-well potential for
$-16/3<h<16/3$.  
$4h$ is the level difference between two potential minima at 
$x=-\Delta\equiv a$ and $x=\Delta\equiv c$. 
The local potential maximum
between the two minima is at
$x=-3h\Delta/16 \equiv b$, and the potential values are
$[U(a),U(b),U(c)]=[-4(1-h/2),(3h/16)^2(8-3h^2/64),-4(1+h/2)]$.
See figure 1. The results are readily generalized to other forms
of the potential $U(x)$.

\begin{figure}[htb]
\epsfxsize=8cm
\epsffile{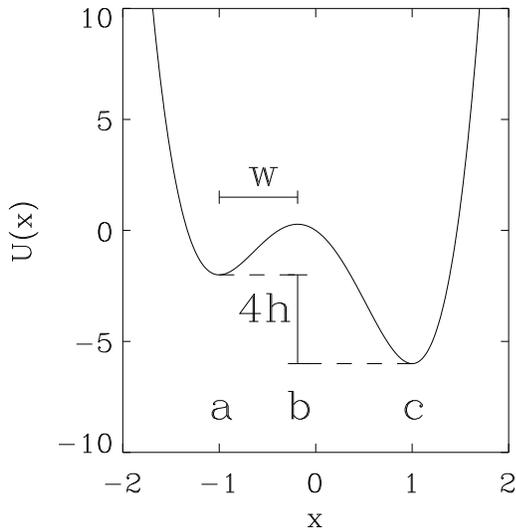}
\caption[]{ 
The potential (\ref{potential}). $4h = U(a)-U(c)$ is the
potential difference between the two minima, $w = b-a$ is the 'left half-width'. Units are arbitrary.
}\end{figure}

\section{Waiting time}
The
waiting time for jumping between the two potential minima (from $a$ to $c$)
of $U(x)$ defined above 
is exponentially distributed. With $p_{ac}(\tau>t)$ being
the probability of staying in minimum $a$ longer than $t$ we have $p_{ac}(\tau>t)=
\exp(-t/T_{ac})$ with a mean waiting time $T_{ac}$.
This follows from the Markov property of the Langevin equation in the 
discrete state limit, since
we have 

\begin{eqnarray}
P(t<\tau<t+\Delta t)/\Delta t = (1-\lambda_{ac}\Delta t)^{t/\Delta t}\lambda_{ac}
\nonumber \\
\rightarrow \lambda_{ac}\exp(-\lambda_{ac}t),
\label{expT}
\end{eqnarray}
as $\Delta t \rightarrow 0$, where $\lambda_{ac}=1/T_{ac}$ is the transition probability intensity. 
In the
non-discrete case, a little more rigorous treatment is needed \cite{Gardiner}. However,
the result holds, if the potential wells are substituted for the minima,
and the waiting time is defined as the time between consecutive 
crossings of $a$ and $c$.  

\subsection{Gaussian noise and Arrhenius formula}
In the case of Gaussian noise in (\ref{LE}) $T_{ac}$
can be calculated from the backward Fokker-Planck equation \cite{Gardiner}, 

\begin{equation}
T_{ac} \approx \frac{2}{\sigma^2}\int_{-\infty}^b dx 
e^{-2U(x)/\sigma^2}\int_a^c dye^{2U(y)/\sigma^2},
\label{Arrh}
\end{equation}
and correspondingly for $T_{ca}$.
By using the saddle-point approximation on (\ref{Arrh}) we obtain the 
Arrhenius formula, 

\begin{equation}
T_{ac} \propto \exp(2[U(b)-U(a)]/\sigma^2). 
\end{equation}
For comparison with the case of $\alpha$-stable noise 
figure 2 displays the standard result of a numerical simulation
in the case of Gaussian noise.
Figure 2 (a) shows the simulated process with the potential in figure 1. 
2 (b) shows the simulated probability density function and the right-hand side of 
(\ref{pdf}). Figure 2 (c) shows the time-scale for jumping as a function 
of the parameter $h$. The time-scale is calculated from the exponential
distribution of times between consecutive crossings of the levels $a$
and $c$. Figure 3 shows the number of crossings (from $a$ to $c$ and
from $c$ to $a$ respectively) with a waiting-time larger than each waiting-time
measured, normalized by the total number of crossings.
These points are situated on straight lines in the semi-logarithmic plot where 
$T_{ac}$ and $T_{ca}$ are the slopes
of the lines. Figure 2 (c) shows the time-scales for seven simulations
with different $h$. The curves are the time-scales
calculated from (\ref{Arrh}).

\begin{figure}[htb]
\epsfxsize=14cm
\epsffile{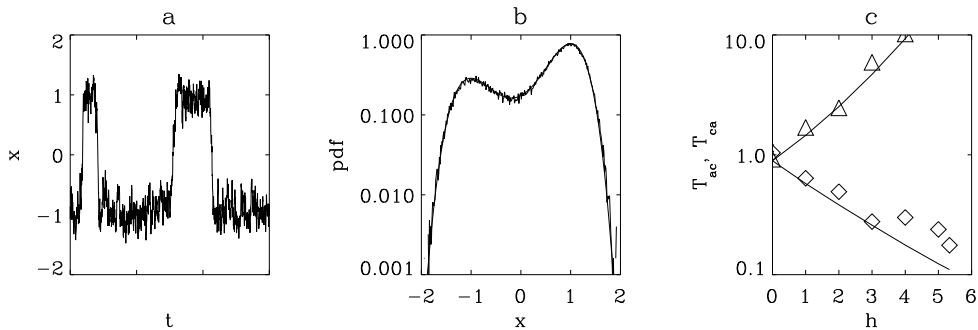}
\caption[]{ 
A simulation of (\ref{LE}) with Gaussian noise and the potential
shown in figure 1. (a) shows a realization and (b) the probability density
function. The actual simulation is 1000 times longer than what is shown in (a).
The smooth curve in (b) is the pdf calculated from (\ref{pdf}). (c) shows the mean waiting
times, $T_{ac}$ (diamons) and $T_{ca}$ (triangles), for seven simulations with
varying $h$. The curves are the waiting times calculated from (\ref{Arrh}). Units are arbitrary.
}\end{figure}

\begin{figure}[htb]
\epsfxsize=14cm
\epsffile{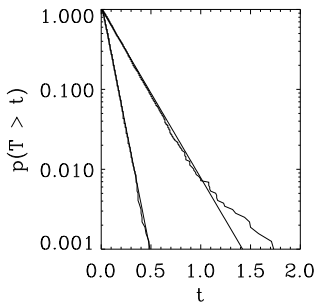}
\caption[]{ 
The probability for waiting longer than $t$ before jumping to the
other well as a function
of $t$ obtained from the simulation. The slope of the upper curve gives
$T_{ca}$ and the slope of the lower curve gives $T_{ac}$. Units are arbitrary.
}\end{figure}

\subsection{$\alpha$-stable noise}
In the case $\alpha < 2$ the
situation is radically different. The sample curves of the process are no longer
continuous and the finite jumps or extreme events 
will contribute to the
probability of jumping between the potential wells. The probability, 
$(\lambda_{ac}\Delta t + o(\Delta t))$, for jumping from the left well,
$x < b$ to
any $y > b$ in a single jump in a time interval $\Delta t$ is
governed by the tail of the distribution, $p(x)\propto 
(x/\sigma)^{-(\alpha+1)}\Delta t/\sigma$. This is seen by observing that
the process (\ref{LE}) can be obtained from the discrete process,
$X(t+\Delta t)=X(t)+f(X(t))\Delta t + [\sigma \Delta t^{1/\alpha}]\eta(t)$,
for $\Delta t \rightarrow 0$, where $\eta(t)$ has an $\alpha$-stable 
distribution with unit intensity. Thus we have

\begin{eqnarray}
\lambda_{ac}\Delta t\approx P(X(t+\Delta t)>b|X(t)<b)/P(X(t)<b) \propto \nonumber \\ 
\int_{-\infty}^b[\int_{b-x}^\infty p(u)du]\tilde{p}(x)dx \approx 
\int^{\infty}_{b-a}p(u)du 
\approx 
[(b-a)/\sigma]^{-\alpha}\Delta t 
\end{eqnarray}
The inner integral is the probability of jumping from $x<b$
to any $y>b$, and $\tilde{p}(x)$ is the stationary
probability density.
The outer integral is dominated by the central part of 
the probability distribution. This result is exact
in the $\Delta t \rightarrow 0, \sigma \rightarrow 0$
limit where $p(x)\rightarrow \delta (x-a)$.
Thus we have
 
\begin{equation}
T_{ac}=c(\alpha)[(b-a)/\sigma]^{\alpha}
\label{tjump}
\end{equation}
where $c(\alpha)$ is some constant.
So in this case we see that the waiting time scales with
the 'left half-width' of the barrier, $b-a \equiv w$, to the 
power $\alpha$. The
height of the barrier has no influence on the transition probability.
The results are confirmed by numerical simulation.
Figure 4 displays the numerical simulation using 
Cauchy noise, $\alpha=1$, and the same potential as in the case displayed in
figure 2. 
Note the linear scale in figure 4 (c) showing the scaling of
the time-scale
with $w$.

\begin{figure}[htb]
\epsfxsize=14cm
\epsffile{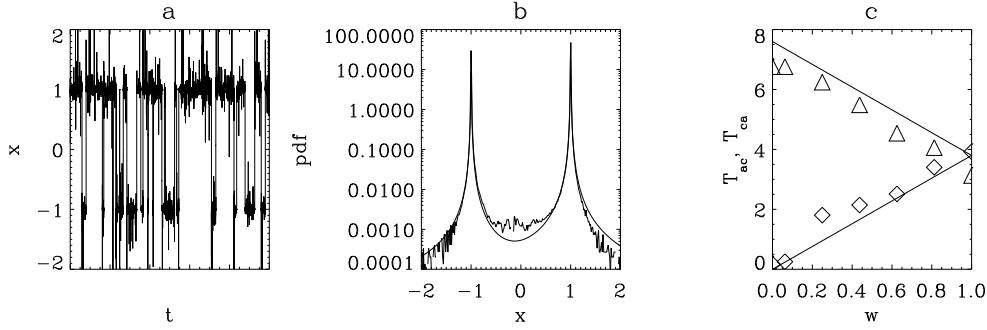}
\caption[]{ 
The same as figure 2 but with Cauchy noise, $\alpha=1$. Note
the linear axis in (c). The curves are obtained from (\ref{tjump}).
}\end{figure}

\section{The stationary distribution}
For the Cauchy noise driven system the indeterminacy in (\ref{8})
can now be resolved by use of the Master equation (\ref{Master}). 
In the limit $\sigma \rightarrow 0$ the system can be
approximated as a discrete two state system, with the two states
corresponding to the two potential minima, at $a$ and $c$. In this limit the
system  fulfill the 
stationary Master equation,

\begin{equation}
0=p_a p(a \rightarrow c) - p_c p(c \rightarrow a).
\end{equation}

The transition probabilities are now $p(a \rightarrow c) \propto 1/T_{ac} \propto (b-a)/\sigma$
and $p(c \rightarrow a) \propto (c-b)/\sigma$ and we get,

\begin{equation}
p_a=1-p_c=(b-a)/(c-a). 
\label{pa}
\end{equation}
Note that this is independent of 
$\exp(-2[U(a)-U(c)]/\sigma^2)$,
which in the Gaussian case corresponds to the Gibbs distribution.
Figure 5 shows the distribution $p_a$, which is different from
the Gibbs distribution, as a function of $w$.
Figure 4 (b) shows the probability density function from the simulation
over-plotted the one calculated from (\ref{8}) and (\ref{pa}).

\begin{figure}[htb]
\epsfxsize=14cm
\epsffile{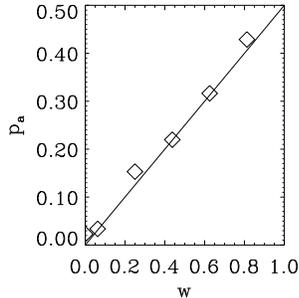}
\caption[]{ 
The probability, $p_a=1-p_c$, for finding the particle
in the left well as a function of $w$ in the simulation with Cauchy noise.
The curve is obtained using (\ref{pa}), The distribution is deviating 
strongly from a 
Gibbs distribution.
}\end{figure}

\section{Barrier penetration}
When $\alpha$ is close to 2 we should expect the 'single jump
penetration' of the barrier to become more and more unlikely and the 
continuous penetration dominating. The Levy decomposition
theorem \cite{Protter} states that the $\alpha$-stable process
can be decomposed in a Brownian process and a compound Poisson
process. The 'continuous' barrier penetration can be
estimated by considering the distribution 
to be truncated so that there are no jumps larger than the
half-width of the barrier, $w$. 
The truncated probability for the noise, $p^t(x)$, is then defined
by  $p^t(x) \propto p(x)$ for $|x| < w$ and
$p^t(x) =0$ for $|x| \ge w$. This part of the noise now has finite
second order moment and we can estimate the variance as 
$\sigma^2_{\mbox{\small eff}}\propto \int x^2 \tilde{p}(x)dx \propto w^{2-\alpha}$ asymptotically for large $w$ or small noise intensity $\sigma$.
The waiting time can be estimated as,

\begin{equation}
T^c\propto \sigma^{-2}_{\mbox{\small eff}}
\exp(2[U(b)-U(a)]/\sigma^2_{\mbox{\small eff}}),
\end{equation}
 where $c$ denotes
'continuous'.
Note that this part of the process is not strictly continuous,
since it contains jumps smaller than $w$.
The time-scale for single jump penetration can be estimated from
(\ref{tjump}),

\begin{equation}
T^d \propto w^\alpha,
\end{equation}
where $d$ denotes 'discontinuous' 
and we have,

\begin{equation}
\frac{T^d}{T^c}\propto
w^2\exp(-\tilde{c} [U(b)-U(a)] w^{\alpha-2}),
\label{tt}
\end{equation}
where $\tilde{c}$ is a constant.
So the relative importance of extremal jumping depends both on the height
and the width of the barrier. To illustrate the relative importance of 
two jumping processes a simulation of (\ref{LE}) with an $\alpha$-stable
noise \cite{Janicki} with $\alpha=1.7$ and a potential 
(\ref{potential}) with $h=3$,
was performed. Figure 6 shows part of a realization of this
process. Here it is seen that the jumping from
the deep to the shallow well is governed by the discontinuous part,
$T^d(c\rightarrow a) \ll 
T^c(c\rightarrow a)$, while the jumping from
the shallow to the deep well is dominated by the 'continuous' part,
$T^d(a\rightarrow c) \gg
T^c(a \rightarrow c)$.
For proportioning the continuous and discontinuous processes in a
given situation the prefactor and the constant, $\tilde{c}$, in (\ref{tt}) must be calculated or
estimated. 

\begin{figure}[htb]
\epsfxsize=12cm
\epsffile{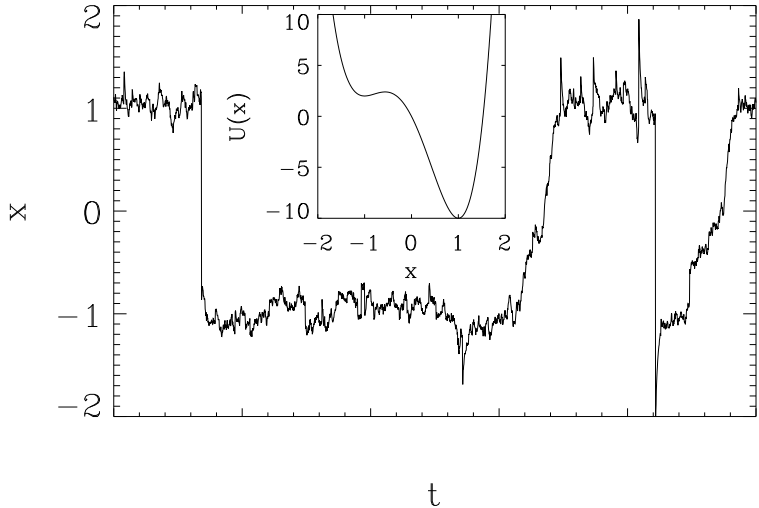}
\caption[]{ 
A realization of the process with $\alpha=1.7$. 
The potential used is shown in the insert. The jumping from the
left (shallow) well to the right (deep) well is triggered by the 
- almost - normal diffusion. The jumping the other way is driven by the 
tail of the $\alpha$-stable distribution, the extreme events. 
}\end{figure}

\section{Summary}
We have seen that the statistics of noise induced 
jumping between
meta-stable states in a potential is different for $\alpha$-stable 
noise from the usual Gaussian noise case. The stationary probability 
distribution deviates from the Gibbs distribution, and the waiting
time for jumping depends in some cases more on the width than on
the height of the barrier. This is the case where a single extreme
event triggers the jumping. These observations might be of importance 
for understanding the triggering mechanisms of climatic changes, where
the flow state of the ocean is trapped in a potential minimum, a stable
climatic state. This flow is stochastically forced by the atmospheric
flow. There are some evidence that this stochastic forcing is $\alpha$-stable
rather than Gaussian such that climatic shifts from one state to another
could be triggered by single extreme events. This would perhaps explain why the 
climate models at present are not capable of reproducing the 
climatic changes observed in the geological records. The models are
too coarse grained and contains to much diffusive smoothening to allow
for extreme events.

\section{Acknowledgement}
I would like to thank O. Ditlevsen
for valuable discussions. The work was funded
by the Carlsberg foundation.

\appendix
\section{The addition of $\alpha$-stable random variables.}
Textbooks on $\alpha$-stable processes are
now available \cite{S&T,Janicki}, but for those readers not familiar
with the $\alpha$-stable distributions and processes a few notes are
added in the following.

When $\{X_i,\, i=1,...,n\}$ is a series of i.i.d. random variables,
the distribution of the variable $Y=c(n)\sum_{j=1}^n X_j$ can be
determined from the characteristic function,

\begin{eqnarray}
\langle \exp[{ikY}]\rangle = \langle \exp[{ikc(n)\sum_{j=1}^n X_j}]\rangle = \nonumber \\
\langle \Pi_{j=1}^n \exp[{ikc(n) X_j}]\rangle = 
\langle e^{ikc(n)X}\rangle^n.
\label{a1}
\end{eqnarray}
If the distribution for $Y$ is the same as for $X$ the equation (\ref{a1})
for the characteristic function, $f(k) = \langle \exp[{ikY}]\rangle$ is,

\begin{equation}
f(k)=f(c(n)k)^n
\label{a11}
\end{equation}
with the solution

\begin{eqnarray}
f(k)=\exp[{-\sigma^\alpha |k|^\alpha/\alpha}], \label{a2} \\
c(n)=n^{-1/\alpha} \label{a3}.
\end{eqnarray}
The constant, $\sigma^\alpha/\alpha$ is chosen so that it coinsides
with the usual notation in the gaussian case, $\alpha=2$.
Only for $\alpha >0$ does (\ref{a2}) represent a characteristic 
function. It can be shown that the characteristic function 
(\ref{a2}) corresponds
to distributions with power-function tails, 
$P(X>x)\sim x^{-\alpha}$ \cite{Feller,S&T}.
For $\alpha > 2$ the second moment of the distribution
exists and sums of i.i.d. variables converges by the central
limit theorem to the gaussian distribution, $\alpha=2$. 
For $0<\alpha<2$ the distributions has a domain of attraction
in the sense that sums of i.i.d. random variables 
with tail distributions, $P(X>x)\sim x^{-\gamma}$, under rather
general conditions 
converges to an $\alpha$-stable distribution with $\alpha=\gamma$.
This is the generalization of the central limit theorem for
$\alpha$-stable distributions. The proof of this is similar
to the proof of the central limit theorem for the normal distribution. 
It basically 
substitudes a limit, $\tilde{f}(c(n)k)^n \rightarrow f(k)$ for
(\ref{a11}). The proof can be found in Fellers book, pp 574 -- 581  \cite{Feller}.

Now we can intuitively understand the noise term, $dL_\alpha$, in the
Langevin equation (\ref{LE}) as the
continuum limit of addition of small increments,

\begin{equation}
\Delta L_{\alpha}(\Delta t) = \frac{1}{m^{1/\alpha}}\sum_{j=1}^m X(j\Delta t/m)
\end{equation}
where $X(t)$ is a random process with power-function tails, $P(X(t)>x) 
\sim x^{-\alpha}$, and unit intensity. In the limit, $m\rightarrow 
\infty$, $\Delta L_\alpha$ will be
an $\alpha$-stable noise. It follows from (\ref{a3}) that $dL_\alpha =
dt^{1/\alpha}$, which in the gaussian case is the well-known relation,
$dB^2 = dt$. 

For $\alpha<2$ the $\alpha$-stable variables have infinite variance. This
concept can be difficult to comprehend when considering measurements from a given physical system. In
the case a sample is taken, say of $n$ measurements of the variable $X$,
where $X$ has an $\alpha$-stable distribution with stability index $\alpha$,
then of course any of the measurements, $x_1,...,x_n$ of $X$ is finite
so that the sample variance, $(x_1^2+ ... +x_n^2)/n$, is some finite
number. The variable $Y=X^2$ will have a tail distribution given by
$P(Y > x^2) = P(X > x) \sim x^{-\alpha} = y^{-\alpha/2}$,
so that, asymptotically for large $n$, $Z_n \equiv n^{-2/\alpha}(Y_1+...+Y_n)$ will have an $\alpha$-stable
distribution with stability index $\alpha/2$. Imagine now
that we estimate the (infinite) variance of variable $X$ by taking samples
of length $n$, estimating the variance as $(X_1^2+ ... + X_n^2)/n=n^{2/\alpha-1}
Z_n$. Then the estimate itself will be an $\alpha$-stable process with
stability index $\alpha/2$ and intensity $n^{2/\alpha-1}$. This estimate will
be fluctuating with an intensity growing with $n$ for $\alpha < 2$.

\section{The Fokker-Planck equation}
In the following the Fokker-Planck equation (\ref{FP}) corresponding
to the Langevin equation (\ref{LE}) will be derived.
The Fokker-Planck equation will be derived in spectral form using 
that the $\alpha$-stable 
processes are defined by their characteristics 
functions. Following the lines of Stratonovich \cite{Stratonovich1963}
we define 
the functional 
\begin{equation}
I = \int R(y) \partial_t p(x_0|y,t) dy = 
\lim_{\Delta t \rightarrow 0}I_{\Delta t }
\end{equation}
where
\begin{equation} 
I_{\Delta t}=\frac{1}{\Delta t}\int R(y) [ p(x_0|y,t+\Delta t) - 
p(x_o|y,t)] dy
\label{int1}
\end{equation}
 $R(y)$ is an arbitrary generator function, and 
$ p(x_0|x_1,t)$ is the conditional probability density at $x_1$ 
corresponding to passing from $x_0$ to $x_1$ 
during time $t$. Assuming 
stationarity we suppress the first temporal index, 
$ p(x_0|x_1,t)\equiv p(x_0,0|x_1,t)= p(x_0,\tau|x_1,\tau+t)$. 
For simplicity of writing we make the convention
that $\int$ is to  be read as
$(1/\sqrt{2\pi})\int_{-\infty}^{\infty}$.

With $ p(x_0|x,t)$ being a probability density in $x$ we trivially have
\begin{equation} 
\int_{-\infty}^{\infty}p(x_0|x,t) dx =1
\end{equation}
and the Chapman-Kolmogorov equation
\begin{equation}
p(x_0|x_1,t) = \int_{-\infty}^{\infty} p(x_0|x,\tau) p(x|x_1,t-\tau) dx. 
\end{equation}
 For the functional we then get
\begin{eqnarray}
I_{\Delta t} = \nonumber \\
\frac{1}{\Delta t}\int R(y) 
[\int_{-\infty}^{\infty}p(x|y,\Delta t) p(x_0|x,t) dx - p(x_0|y, t)]dy 
 \nonumber \\
=\frac{\sqrt{2\pi}}{\Delta t}\int p(x_0|x,t)\{\int p(x|y,\Delta t)
[R(y)-R(x)] dy\}dx
\end{eqnarray}
 We now define the Fourier transforms 
\begin{equation}
R(x)=\int \hat{ R}(k)e^{ikx}dk,\: \hat{ R}(k)=\int R(x)e^{-ikx}dx,
\end{equation}
similarly for $f(x)$ in (\ref{LE}), and for $\sigma^\alpha(x)$ 
to be introduced below. However, for the probability density 
$p(x_0|x,t)$ we define 
\begin{eqnarray}
p(x_0|x,t)=\int \hat{p}(x_0|k,t)e^{-ikx}dk, \\ 
\hat{p}(x_0|k,t)=\int p(x_0|x,t)e^{ikx}dx
\end{eqnarray}
consistent with the standard definition of characteristic function
except for the factor
$\sqrt{2\pi}$. 
With these definitions it is easy to derive the formula
\begin{equation}
\int f(x)p(x_0|x,t) e^{ikx}dx=\int\hat{f}(k_1-k) \hat{p}(x_0|k_1,t) dk_1 
\label{conv}
\end{equation}
from which it directly follows that
\begin{equation}
 f(x)p(x_0|x,t)=\int[\int\hat{f}(k_1-k) \hat{p}(x_0|k_1,t) dk_1]
e^{-ikx}dk 
\label{invconv}
\end{equation}
Using the spectral representation
for the generator function we get
\begin{eqnarray}
\frac{1}{\sqrt{2\pi}}I_{\Delta t}
=\int p(x_0|x,t)\{\int p(x|y,\Delta t)\frac{1}{\Delta t}
\int\hat{R}(k)(e^{iky}-e^{ikx}) dk\}dy \nonumber \\
=\int\int e^{ikx} \hat{R}(k)  p(x_0|x,t) \frac{1}{\Delta t}
E[e^{ik[X(t+\Delta t)-x]}-1|X(t)=x] dk dx
\label{int2}
\end{eqnarray}
The conditional expectation 
is evaluated using the Langevin equation (\ref{LE}) and 
the characteristic function of the of the alpha-stable Levy noise
increment
$dL_\alpha$, 
\begin{equation}
E\{\exp[ikdL_\alpha]\}=\exp[-\sigma^\alpha \Delta t
|k|^\alpha/\alpha]
\label{charLevy}
\end{equation}
We get
\begin{eqnarray}
 \frac{1}{\Delta t}
E[e^{ik[X(t+\Delta t)-x]}-1|X(t)=x]
\nonumber \\
=\frac{1}{\Delta t}
E[e^{ik[f(x)\Delta t+o(\Delta t)+dL_\alpha]}-1]
\nonumber \\
= \frac{1}{\Delta t}
e^{ik[f(x)\Delta t+o(\Delta t)]
-\sigma^\alpha \Delta t |k|^\alpha/\alpha}-1
\nonumber \\ 
\rightarrow ikf(x)
-\sigma^\alpha |k|^\alpha/\alpha
\label{expectation}
\end{eqnarray}
as $\Delta t \rightarrow 0$.
Substitution of (\ref{expectation}) in (\ref{int2}) and combining 
with (\ref{int1}) then gives
\begin{eqnarray}
\int\int \hat{R}(k)e^{ikx} \partial_t p(x_0|x,t) dx dk
 \nonumber \\
=\int\int\hat{R}(k)e^{ikx} [ikf(x)- 
\sigma^{\alpha}(x)|k|^{\alpha}/\alpha] \, p(x_0|x,t) dx dk
\end{eqnarray}
Here we have permitted the scaling factor $\sigma^{\alpha}/\alpha$,
 corresponding to the variance  $\sigma^2 dt$ of the noise increment in
the case 
$\alpha =2$, to depend on the variable $x$. 
By eliminating the $x$ by use of (\ref{conv})
we get, suppressing $x_0$,
\begin{equation}
\int \hat{ R}(k)\partial_t \hat{p}(k,t)dk
 = \int\int \hat{ R}(k)[ik \hat{f}(k_1-k)-
\hat{\sigma^{\alpha}}(k_1-k) |k|^{\alpha}/\alpha) \hat{p}(k_1,t)] dk dk_1
\end{equation}
and finally since $\hat{R}(k)$ is arbitrary we get the spectral 
Fokker-Planck equation for the integrand, suppressing the $t$ 
index,
\begin{equation}
\partial_t \hat{p}(k) 
= \int (ik \hat{f}(k_1-k)-
\hat{\sigma^{\alpha}}(k_1-k)
|k|^{\alpha}/\alpha)\hat{p}(k_1)dk_1
\label{spectralFP}
\end{equation}
Multiplying by $e^{-ikx}$ and using (\ref{invconv}) gives the 
Fokker-Planck equation in the usual form
\begin{equation}
\partial_t p(x) = -\partial_x
[f(x)p(x)]
-\frac{1}{\alpha}\int\int e^{-ikx} \hat{\sigma^{\alpha}}(k-k_1) 
|k|^{\alpha} \hat{ p}(k_1) dk dk_1
\label{FP1}
\end{equation}
For the stationary Fokker-Planck equation the l.h.s of (\ref{FP1})
vanishes and 
the partial derivatives become total derivatives.
The last term on the r.h.s 
is a generalized diffusion which 
formally can be written
\begin{eqnarray}
\frac{1}{\alpha}\frac{d^{\alpha}}{dx^{\alpha}}[\sigma^{\alpha}(x)p(x)]
\nonumber \\ 
\equiv -\frac{1}{\alpha}\int\int e^{-ikx} \hat{\sigma^{\alpha}}(k-k_1) 
|k|^{\alpha} \hat{p}(k_1) dk dk_1
\end{eqnarray}
In the case $\alpha = 2$ this is the usual diffusion 
term corresponding to Gaussian white noise excitation of intensity
$\sigma^2(x)$. 
For $\alpha < 2$ the diffusion  is non-local. The physical meaning of 
this term is that for $\alpha$-stable processes there will, due to the fat 
tails of the distributions, be finite size jumps in the 
process.

\section{The Cauchy distribution}
The probability density can only be expressed explicitely for $\alpha=1$, and $\alpha=1/2$.
For $\alpha=1$, Cauchy noise, the
characteristic function is $c(k)=\exp(-\sigma |k|)$ and its Fourier transform is,

\begin{equation}
p(x)=\frac{1}{\pi \sigma [1+(x/\sigma )^2]}
\label{b1}
\end{equation}
For this distribution even the mean does not exist. Note that even though
the density distribution (\ref{b1}) is symmetric, $p(-x)=p(x)$, this
does not imply that $\langle x\rangle =0$. For a data sampling
this manifests itself in the fact that the average of $n$ data points,
$Z_n=(X_1+ ... +X_n)/n$ is Cauchy distributed with the same intensity as $X_i$,
so that there is no convergence for the series $Z_n, \, n=1,2,...$, it fluctuates
exactly as the data $X_i$ itself.

A classical example
of this characteristic of the Cauchy distribution is seen by considering
the distribution of light on a line, $L$, from a point source,
see figure 7. Since the light is uniformly distributed over the angles,
$p(\theta)=1/\pi, \, \theta \in [0,\pi]$, the distribution on the
line will be, $p(x)=p(\theta)(d\theta /dx)=1/[\delta \pi(1+(x/\delta)^2]$, where $X=\delta \tan(\theta)$ is a stochastic variable representing the point
where a foton released from $S$ at the (stochastic) angle $\theta$ crosses $L$.
Now inserting $n-1$ lines, $L_i$, 
parallel to $L$, between the light source, $S$, and $L$, we
can apply Huygens principle, saying that $L_i$ will act as a line of
point sources, where the light follows the path $S \rightarrow X_1 \rightarrow 
X_1+X_2\rightarrow ... \rightarrow X$, where $X=X_1+ ... + X_n$. The variables
$X_i$ are independent and Cauchy distributed with scale parameter $\delta/n$.
Thus Huygens principle is consistent with the fact that $X=(\sum_{i=1}^n X_i)/n$ has the same distribution as $X_i$.

\begin{figure}[htb]
\epsfxsize=12cm
\epsffile{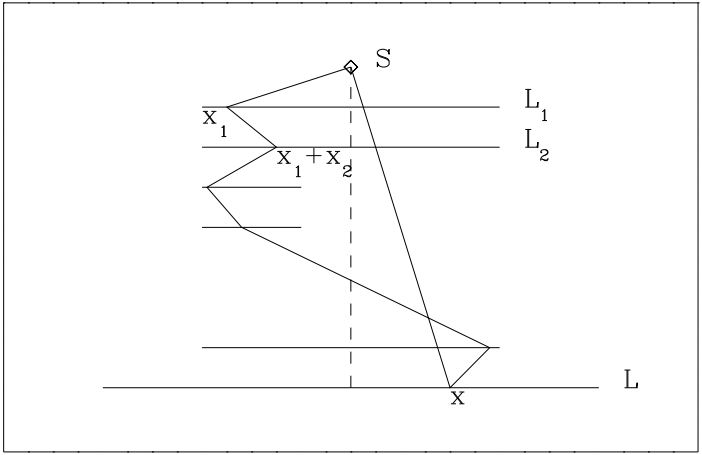}
\caption[]{ 
Huygens principle applied to the light from a point source on a
line. This illustrates the behavior of the averaging of Cauchy distributed stochastic variables.
}\end{figure}

\section{Simulations}
The simulations performed in this work only involves Cauchy noise, which
is easily obtained from a random variable, $X$, uniformly distributed 
in the interval $[-\pi/2,\pi/2]$, as
$Y=\tan(X)$. More generally a random variable
with an $\alpha$-stable distribution \cite{Janicki}
is obtained from,

\begin{equation}
Y=[\sin(\alpha X)/\cos(X)^{(1/\alpha )}]\times
[-\cos([1-\alpha ]X)/\log(W)]^{(1-\alpha )/\alpha}
\label{d1}
\end{equation}
where $X$ is defined as above and $W$ is another random 
variable uniformly distributed on the interval $[0,1]$.
When simulating (\ref{LE}) by a discrete numerical time stepping
the (fixed size) time steps usually needs to be much smaller
than would be expected from numerical integration of the 
drift term alone. This is due to the large excursions from the
tails of the distribution of the noise. It is thus important
to use a stable integration routine for the drift term. A
simple durable routine, which is the one used in these simulations,
is Heun's integration scheme. The simulation is performed as 
$x(t+\Delta t)=x(t)+(f[x(t)]+f[x(t)+
f[x(t)\Delta t])\Delta t/2 +\sigma \Delta t{1/\alpha}\eta(t)$,
where $\eta(t)$ is generated by (\ref{d1}).



\end{document}